\documentclass{LMCS}
\usepackage{amsmath,amssymb,bbm,hyperref}

\newcommand{\tmem}[1]{{\em #1\/}}
\newtheorem{definition}{Definition}
\newcommand{\assign}{:=}
\newtheorem{theorem}{Theorem}
\newcommand{\emdash}{---}
\newtheorem{corollary}{Corollary}
\newtheorem{lemma}{Lemma}

\newcommand{\Z}{\ensuremath{\mathbbm{Z}}}
\newcommand{\C}{\ensuremath{\mathbbm{C}}}
\newcommand{\R}{\ensuremath{\mathbbm{R}}}

\newcommand{\Span}{\ensuremath{\mathrm{span}}}

\def\doi{1 (3:3) 2005}
\lmcsheading%
{\doi}
{355--362}
{}
{}
{Jul.~\phantom{0}6, 2005}
{Dec.~20, 2005}
{}

\begin{document}

\title{Almost periodic functions, constructively}
 \author[B.~Spitters]{Bas Spitters\rsuper a}
\thanks{{\lsuper a}The author was supported by the Netherlands Organization for Scientific  Research (NWO)}
\address{Institute for Computing and Information Sciences\\
Radboud University of Nijmegen}
\email{spitters@cs.ru.nl}
 
\begin{abstract}
  Almost periodic functions form a natural example of a non-separable
  normed space. As such, it has been a challenge for constructive
  mathematicians to find a natural treatment of them. Here we present
  a simple proof of Bohr's fundamental theorem for almost periodic
  functions which we then generalize to almost periodic functions on
  general topological groups.
\end{abstract}
\maketitle

\section{Introduction}

Finding a natural constructive treatment of the theory of almost periodic
functions has long been a challenge for constructive mathematics,
see~{\cite{Brom:1977}}, {\cite{Gibson:1972}} and~{\cite{Margenstern:1978}}.
The present approach is similar to the one by Bishop's student
Brom~{\cite{Brom:1977}}. However, we replace his long and explicit
construction by a simple definition of a new metric on the group, due to von
Neumann~{\cite{vNeumann:ap}}(p.447), and applying Fourier and Peter-Weyl
theory. Thus we obtain a very similar, but much more conceptual construction.

By constructive mathematics we mean constructive in the sense of
Bishop~{\cite{Bishop/Bridges:1985}}. That is, using intuitionistic logic and
an appropriate set theory, or type theory. Like Bishop, we will freely use the
axiom of (countable) dependent choice.

To introduce the theory of almost periodic functions we first consider a
{\tmem{periodic}} function $f : \R \to \C$ with period $2 \pi$, say. We may
identify $f$ with a function $g$ on the unit circle, by defining $g ( e^{ix} )
: = f ( x ),$ for all $x \in \R$. Because the circle is a compact Abelian
group, Fourier theory may be used to approximate the periodic function $f$ by
finite sums of characters.

The sum{\footnote{We use the notation $\lambda x.t$ for the function that
returns $t$ on input $x$. An alternative notation would by $x \mapsto t$.}}
$\lambda x.e^{ix} + e^{i 2 x}$ is periodic, and this remains true when 2 is
replaced by any other rational number. However, the sum of the periodic
functions $\lambda x.e^{ix}$ and $\lambda x.e^{i \sqrt{2} x}$ is not periodic.
But this sum \textit{is} almost periodic. A real function $f$ is {\tmem{almost
periodic}} \label{apBohr} if for every $\varepsilon > 0$, there exists $l > 0$
such that every interval $[ t, t + l ]$ contains at least one number $y$ for
which $|f ( x ) - f ( x + y ) | < \varepsilon$ whenever $x$ is in
$\mathbbm{R}$. Classically, the class of almost periodic functions is closed
under addition and multiplication, but constructively this is not the case.
When $a \neq 0$ the function $\lambda x.e^{iax}$ is periodic. When $a
\downarrow 0$ the period tends to $\infty$. Consequently, when we do not know
whether $a = 0$ or $a \neq 0$, we are unable to show that $\lambda x.e^{iax}$
is almost periodic. In~{\cite{Gibson:1972}} it is shown that the function
$\lambda x.e^{iax}$ is almost periodic if and only if $a = 0$ or $a \neq 0.$
Observe that if $a$ is close to 0, then both $e^{i ( 1 + a ) x}$ and $e^{-
ix}$ are almost periodic. However $e^{i ( 1 + a ) x} e^{- ix} = e^{iax}$ is
almost periodic if and only if $a = 0$ or $a \neq 0.$ Consequently, one can
not prove constructively that the almost periodic functions are closed under
multiplication. A similar argument shows that the almost periodic functions
are not constructively closed under addition. It is, however, straightforward
to prove that the almost periodic functions are closed under uniform limits.

\section{Preliminaries}

In this section we collect some results from constructive topological group
theory, mostly following~{\cite{Bishop/Bridges:1985}}
{\cite{CoquandSpittersPW}}. We used~{\cite{Loomis:AHA}},
{\cite{Sternberg:group}} and~{\cite{Naimark/Stern:GroupRep}} as general
references for the classical theory.  In this section $G$ will denote a
topological group.

\begin{definition}
  A {\tmem{topological group}} is a group that is also a topological space in
  such a way that the group operations are continuous.
\end{definition}

Any locally compact group allows a unique translation invariant measure,
called Haar measure~{\cite{Bishop/Bridges:1985}}. For a simple construction of
Haar measure on compact groups see~{\cite{CoquandSpittersPW}}. This
construction follows von Neumann's classical existence proof.

Let $G$ be a compact group. Let $L_1$ denote $L_1 ( G )$, the space of
Haar-integrable functions on $G$. Define the \textit{convolution} from $L_1
\times L_1$ to $L_1$ by
\begin{equation}
  f \ast g = \lambda x. \int f ( y ) g ( y^{- 1} x ) dy,
  \label{form:convolution}
\end{equation}
for all $f, g \in L_1$. This map is continuous, in fact $\| f \ast g \|_1 \leq
\| f \|_1 \| g \|_1 .$ Define the involution $\tilde{f} \assign \lambda x.f (
x^{- 1} )$. With this involution and convolution as multiplication $L_1$ is a
*-algebra, called the \textit{group algebra}. The group algebra contains much
information about the group. For instance a group is Abelian if and only if
its group algebra is Abelian. The group $G$ is compact, so its Haar measure is
finite, consequently $L_2 ( G ) \subset L_1 ( G )$, and thus, the convolution
product $f \ast g$ belongs to $L_2$, for all $f \in L_1$ and $g \in L_2$. 
Moreover, $\| f \ast g \|_2 \leq \| f \|_1 \| g \|_2 .$ Thus an element $f$ of
$L_1$ can be considered as an operator $\lambda g.f \ast g$ on the Hilbert
space $L_2$. These operators are compact, and thus normable, so $L_1$ can thus
be completed to a C*-algebra. This allows us to use:

\begin{theorem}
  \textup{[Gelfand]} Let $\mathcal{A}$ be a unital commutative C*-algebra. The
  spectrum $X$ of $\mathcal{A}$ {\emdash} that is, set of C*-algebra morphisms
  from $\mathcal{A}$ to $\C$ {\emdash} can be equipped with a topology such
  that $X$ is compact and the Gelfand transform $\hat{\cdot} : \mathcal{A} \to
  C ( X )$, defined by $\hat{a} ( x ) : = x ( a ),$ is a C*-isomorphism.
\end{theorem}

We will sometimes speak about the spectrum of a *-algebra when we mean the
spectrum of the C*-algebra as constructed above.

Let $Z$ denote the center of the group algebra {\emdash} that is the set of
$f$ such that $f \ast g = g \ast f$ whenever $g$ is in $L_1$. A
\textit{character} is a C*-algebra morphism from $Z$ to $\mathbbm{C}$. In case
the group is Abelian, $Z = L_1$ and the characters are in one-one
correspondence~{\cite{Bishop/Bridges:1985}}(p.425) with continuous functions
$\alpha : G \to \C$ such that $| \alpha ( x ) | = 1$ and $\alpha ( xy ) =
\alpha ( x ) \alpha ( y ),$ for all $x, y \in G.$ Remaining in the Abelian
case, the characters, with the usual multiplication of functions, form a
group, denoted $G^{\ast}$. This group is called the \textit{dual group} or
\textit{character group}. We equip the character group with the metric induced
by the sup-norm $\| \cdot \|_{\infty}$.

\begin{theorem}
  \textup{{\cite{Bishop/Bridges:1985}} (Thm. 8.3.17)} The character group
  $G^{\ast}$ of a locally compact Abelian group $G$ is a locally compact
  Abelian group.
\end{theorem}

Following~{\cite{Bishop/Bridges:1985}} we define `(locally) compact space' to
mean (locally) compact metric space. In fact, considering the more general
case of a locally compact group, Bishop and Bridges introduce a new metric
$\rho^{\ast}$ on $G^{\ast}$ such that $( G^{\ast}, \rho^{\ast} )$ is locally
compact. For compact groups this metric is equivalent to the metric induced by
the norm $\| \cdot \|_{\infty}$ on $C ( G ),$ see {\cite{Bishop/Bridges:1985}}
(Lemma 8.3.16).

As a paradigm consider the Abelian group $G : = ( \{ e^{it} : t \in \R \},
\cdot ) .$ The character group of this group is the space of the functions $\{
\lambda z.z^n : n \in \Z \} \subset C ( G )$ with the metric and
multiplication inherited from the normed space $C ( G )$ of continuous
functions on $G$. In this case $G$ is compact and $G^{\ast}$ is discrete. This
is the general situation.

\begin{theorem}
  \label{dualdiscrete}Let $G$ be a compact Abelian group. Then the character
  group $G^{\ast}$ is discrete.
\end{theorem}

\begin{proof}
  Let 1 be the constant function with value 1. This function is a character.
  The set of characters $\alpha$ with $\| \alpha - 1 \|_{\infty} < 1$ is an
  open set which contains only the character 1. Indeed, if $\alpha ( x ) \neq
  1$ for some $x \in G,$ then for some $n,$ $\alpha ( x )^n \in \{ z : \Re z
  \leq 0 \} .$ Since $\alpha ( x )^n = \alpha ( x^n )$, it follows that $|
  \alpha ( x^n ) - 1| \geq 1$. Consequently, there is a neighborhood of the
  character 1 which contains only this character. By translation of this
  neighborhood one obtains for each element in the group $G^{\ast}$ a
  neighborhood containing only that element.
\end{proof}

Every inhabited discrete separable metric space is countable, since any dense
subset must coincide with the whole space. If $G$ is a compact group, then
$G^{\ast}$ is locally compact, and hence separable, since moreover, $G^{\ast}$
is discrete, it is countable.

For a locally compact space $X$ let $C_{\infty} ( X )$ denote the set of
functions that are `zero at infinity'. Bishop and
Bridges~{\cite{Bishop/Bridges:1985}} (p.431,p.442) proved the following
Fourier theorem.

\begin{theorem}
  Let $G$ be a locally compact Abelian group. There is a norm-decreasing
  linear map $\mathcal{F}$ from $L_1 ( G )$ to $C_{\infty} ( G^{\ast} )$ such
  that $\mathcal{F} ( f \ast g ) = \mathcal{F} ( f ) \mathcal{F} ( g )$ and
  $\mathcal{F} ( f ) = \lambda \alpha . \int f ( x ) \alpha ( x ) dx$ whenever
  $f, g$ are in $L_1 ( G )$. The map $\mathcal{F}$ is called the Fourier
  transform. Haar measure $\mu^{\ast}$ on $G^{\ast}$ can be normalized in such
  a way that $\mathcal{F}$ preserves the $L_2$-norm on $L_1 ( G ) \cap L_2 ( G
  )$ and the map $\mathcal{F}^{\ast} : L_1 ( G^{\ast} ) \to C_{\infty} ( G )$
  defined by $\mathcal{F}^{\ast} ( \phi ) : = \lambda x. \int \overline{\alpha
  ( x )} \phi ( \alpha ) d \mu^{\ast} ( \alpha )$ has the following property:
  $\mathcal{F}^{\ast} \mathcal{F} f = f$, for all $f \in L_1 ( G ) \cap L_2 (
  G ) .$
\end{theorem}

In~{\cite{CoquandSpittersPW}} the following results were proved for a general,
not necessarily Abelian, compact group $G$. In this context $Z$ denotes the
center of the group algebra and $\Sigma$ denotes a locally compact subset of
its spectrum as a C*-algebra. The points of $\Sigma$ are called characters. We
recall that $\Sigma$ is discrete.

We define the linear functional $I ( f ) : = f ( e )$ on the group algebra and
remark that $f \ast g ( e ) = ( f, \tilde{g} )$, the inner product with
respect to the Haar integral.

\begin{theorem}
  \label{thm:Planch-pos}Let $f$ be an element of $Z$ such that $\hat{f} \ge
  0$. Let  $a_{\sigma} : = \hat{f} ( \sigma ) / \| \chi_{\sigma} \|_2^2$
  whenever $\sigma$ is an element of $\Sigma$. Then $\hat{f} ( \sigma ) =
  a_{\sigma} \widehat{\chi_{\sigma}} ( \sigma ),$ $I ( f ) = \sum a_{\sigma}$
  and $f = \sum a_{\sigma} \chi_{\sigma}$ uniformly.
\end{theorem}

Let $e_{\sigma} : = \chi_{\sigma} / \| \chi_{\sigma} \|_2$ and $b_{\sigma} ( f
) : = ( f, e_{\sigma} ) .$ Then $\| e_{\sigma} \|_2 = 1$ and $b_{\sigma} =
\hat{f} ( \sigma ) / \| \chi_{\sigma} \|_2 .$

\begin{corollary}
  \textit{\textup{[Plancherel]}} For all $f$ in $Z,$ $I ( f \ast \tilde{f} ) =
  \sum |b_{\sigma} |^2$ and $e_{\sigma}$ is an orthonormal basis for the
  pre-Hilbert space $Z$.
\end{corollary}

The main theorem in the Peter-Weyl theory may be formulated as follows.

\begin{theorem}
  \label{pw}For each $f \in C ( G ),$ $\sum_{\sigma} e_{\sigma} \ast f,$ where
  $\sigma \in \Sigma,$ converges to $f$ in $L_2 .$
\end{theorem}

Usually, the Peter-Weyl theorem speaks about irreducible representations.
Fortunately, these representations are in one-one correspondences with the
characters above.

\section{Almost periodic functions on Abelian groups}

In this section we will prove a constructive Bohr approximation theorem for
the Abelian groups.

As is well-known, it is in general not possible to compute the norms of
constructive analogues of non-separable normed spaces. Fortunately, there are,
at least, two solutions to this problem: using quasi-norms, or using
generalized real numbers. We repeat the definition from
{\cite{Bishop/Bridges:1985}} (p.343).

\begin{definition}
  Let $X$ be a linear space over a scalar field $F,$ where either $F = \R$ or
  $F = \C$. A {\tmem{seminorm}} $\| \cdot \|$ is a map from $X$ to
  {\R} such that for all $a \in F$ and $x, y \in X,$ $\| x \| \geq 0,$ $\| ax
  \| = |a| \| x \|$ and $\| x + y \| \leq \| x \| + \| y \| .$ A
  {\tmem{quasinorm}} on $X$ is a family $\{ \| \cdot \|_i : i \in I \}$ of
  seminorms on $X$ such that for each $x \in X,$ the set $\{ \| x \|_i : i \in
  I \}$ is bounded. Define the apartness relation $\neq$ on $X$ by $x \neq y$
  if and only if there exists $i \in I$ such that $\| x - y \|_i > 0.$
  Likewise, define the equality $x = y$ if and only if not $x \neq y,$for all
  $x, y \in X.$ Then $( X, \{ \| \cdot \|_i : i \in I \} )$ is called a
  \textit{quasinormed space}.
\end{definition}

A quasinormed space may also be viewed as a normed space where the norm is a
generalized real number in the sense of Richman~{\cite{richman:cuts}}. That
is, the norm is a Dedekind cut in the real numbers, but this cut does not need
to be located.

Although previous constructive developments considered only almost periodic
functions defined on the real numbers it is natural to consider almost
periodic functions over general groups. One of von Neumann's main ideas is
that the almost periodic functions allow us to mimic most of the important
constructions of compact groups. Most importantly, one can define a `Haar
measure' on the set of almost periodic functions.
See~{\cite{CoquandSpittersPW}} for a constructive proof of von Neumann's
classical `construction' of Haar measure on compact groups.

Here, instead of mimicking these constructions, we take a slightly different
path: we define a new topology on $G$ and use the theory of compact groups
directly. Classically, this would be slightly less general since we exclude
non-continuous functions. Constructively, one can not define (total)
non-continuous functions. So, like Loomis, we restrict ourselves to continuous
functions.

Let $F$ denote either {\R} or {\C} and let $C_b ( X, F )$ denote the bounded
continuous $F$-valued functions on the set $X$. We will drop the field $F$
when it is either clear from the context, or irrelevant. The space $C_b ( X )$
is a quasi-normed space with quasi-norm $\{ \| \cdot \|_x : x \in X \},$ where
$\| f \|_x$ is defined as $|f ( x ) |$.

A subset $A$ of a quasi-normed space is called {\tmem{totally bounded}} if for
each $\varepsilon > 0$, there is a finitely enumerable set $f_1, \ldots, f_n
\in A$ such that for each $f \in A$, there exists $i$ such that $\| f - f_i
\|_j < \varepsilon$ whenever $j$ is in $I$. Note that unlike in the metric
case, we can not require all the elements $f_1, \ldots, f_n$ to be distinct.

Let $G$ be an Abelian group. Define the operator $T_s$ from $C_b ( G )$ to
$C_b ( G )$ by $T_s = \lambda f. \lambda x.f ( s + x )$ for all $s$ in $G$.

\begin{definition}
  A function $f \in C_b ( G )$ is {\tmem{almost periodic}} if the set $S_f : =
  \{ T ( s ) f : s \in G \}$ is a totally bounded subset of $C_b ( G )$.
\end{definition}

In case $G = \R$, this definition is equivalent to Bohr's original definition,
which we stated on p.\ref{apBohr}. Loomis' proof~{\cite{Loomis:AHA}} (41F,
p.171) of this fact is constructive. We note that every almost periodic
function is uniformly continuous.

Let $f$ be an almost periodic function on $G$. The function $v_a : = \lambda
g.g ( a )$ is a uniformly continuous function from $C_b ( G )$ to $G.$ Because
$S_f$ is totally bounded, we may define a pseudometric on $G$ by
\[ d_f ( a, b ) : = \sup_{g \in S_f} |v_a ( g ) - v_b ( g ) | = \sup_{x \in G}
   |f ( a + x ) - f ( b + x ) |. \]
This metric is invariant under the action of the group $( G, + ),$ that is,
$d_f ( a + c, b + c ) = d_f ( a, b ),$ for all $a, b, c \in G.$ Since $\| T_y
f - T_z f \|_{\infty} = \sup_x |f ( x + y ) - f ( x + z ) | = d_f ( y, z )$,
there is an isometric embedding from $S_f$ into $( G, d_f )$. Consequently, $(
G, d_f )$ is totally bounded and we let $G_f$ denote the completion of $( G,
d_f )$, which is a compact group.

Recall from theorem~\ref{dualdiscrete} that the character group of a metric
compact {\tmem{Abelian}} group is discrete. We obtain the following Plancherel
theorem for almost periodic functions.

\begin{theorem}
  \label{Thm: ap Fourier}Let $f$ be an almost periodic function. Let $\Sigma =
  G_f^{\ast}$ be the character group of the compact group $G_f$. Then $f$ is a
  continuous function on $G_f$ and $f = \sum \hat{f} ( \sigma ) \chi ( \sigma
  )$ in $l_2 ( \Sigma ) .$
\end{theorem}

The function $f$ is uniformly continuous. So for every $\varepsilon > 0$,
there is a $\delta > 0$, such that when $|a - b| < \delta$, then $|f ( a + x )
- f ( b + x ) | < \varepsilon$ for all $x \in G.$ Hence $d_f ( a, b ) \leq
\varepsilon .$ Consequently, any continuous function on $G_f$ is a continuous
function on $G$, and thus characters of $G_f$ are characters of $G.$ This
shows that the space $\Sigma$ in the previous theorem is the canonical choice.

The following theorem is called the Bohr approximation theorem. As we remarked
before, in constructive mathematics a sum, and therefore a linear combination,
of characters need not be almost periodic.

\begin{theorem}
  \label{Thm: ap unif R}Let $f$ be an almost periodic function on $G$. Then
  $f$ can be approximated uniformly by an almost periodic linear combination
  of characters.
\end{theorem}

\begin{proof}
  Let $I_G$ be a subset of $\mathbbm{N}$ which is in bijective correspondence
  with $\Sigma$. Then $\Sigma = \{ \chi_n : n \in I_G \}$ and we define $I_n
  \assign \{ i \leq n : i \in I_G \}$. Let $P_n$ be the projection in $L_2$ on
  $\Span \{ \chi_i : i \in I_n \} .$ Then $P_n f \to f$ uniformly as $n
  \rightarrow \infty$, by theorem~4.3~{\cite{CoquandSpittersPW}}. Since $P_n =
  \lambda f. \sum_{i \in I_n} \chi_i \ast f,$ it follows from equation (3.4)
  in~{\cite{CoquandSpittersPW}} that $P_n$ commutes with $T_s$ whenever $s$ is
  in $G$. Consequently, $S_{P_n f} = \{ T_s P_n f : s \in G \} = P_n S_f .$
  This set is totally bounded, because $S_f$ is totally bounded and $P_n$ is
  uniformly continuous. It follows that $P_n f$ is almost periodic.
\end{proof}

The measure $\mu$ we used in Theorem~\ref{Thm: ap Fourier}, Haar measure on
$G_f$, may seem a little ad hoc. In fact $\mu ( f )$ is equal to the value of
the unique constant function in closure of the convex hull of $S_f$. See the
construction of Haar measure in~{\cite{vNeumann:ap}} and its constructive
variant~{\cite{CoquandSpittersPW}}. In the case $G = \R,$ the number $\mu ( f
)$ is also equal to $M ( f ) = \lim_{N \to \infty} \frac{1}{2 N} \int_{- N}^N
f$ which is usually used in this theorem. See~{\cite{Loomis:AHA}} (p.171) for
a constructive proof of this fact. Classically, $M$ is an integral on the
space of almost periodic functions. Constructively, the sum of two almost
periodic functions need not be almost periodic, so $M$ can not be an integral
on the set of all almost periodic functions.

\section{Almost periodic functions on general topological groups}

In this section we extend the results from the previous section to arbitrary
topological groups. Therefore, we let $G$ denote a topological group and $e$
denote its unit.

\begin{definition}
  Let $f$ be a bounded continuous function on $G.$ Define the operators $T_s :
  = \lambda g \lambda x.g ( sx )$ and $T_{}^s : = \lambda g \lambda x.g ( xs
  )$, for all $s \in G$. A function $f$ is \tmem{left almost periodic} if
  the set $S_f = \{ T_s f : s \in G \}$ is totally bounded in $C_b ( G )$, it
  is \tmem{right almost periodic} if the set $S^f = \{ T^s f : s \in G \}$
  is totally bounded in $C_b ( G )$. Finally, $f$ is called \tmem{almost
  periodic} if it is both left and right almost periodic.
\end{definition}

In the following it is often the case that the proof that $S^f$ is totally
bounded is symmetric to the proof that $S_f$ is totally bounded. In such cases
we will only prove the latter statement.

\begin{lemma}
  Every almost periodic function $f$ is normable.
\end{lemma}

\begin{proof}
  If $f$ is almost periodic, then so is $|f|$. Now, $f = \lambda s. ( T_s f )
  ( e ) \text{}$ and hence $\| f \| = \sup_{s \in G} |f ( s ) | = \sup_{g \in
  S_{|f|}} g ( e )$ exists.
\end{proof}

Every continuous function $f$ on a compact group $H$ is almost periodic.
Indeed, $S_f$ is totally bounded, since it is the uniformly continuous image
of the compact set $H$.

Let $f$ be almost periodic on $G.$ Define the pseudo-metric
\[ d_f ( a, b ) : = \sup_{g \in S_f} |g ( a ) - g ( b ) | \]
on $G.$ As before, the space $( G, d_f )$ is totally bounded, and its
completion $G_f$ is a compact group. The function $f$ is continuous on $G_f$,
and by the Peter-Weyl theorem~\ref{pw} $f = \sum_{\chi} f \ast \chi$ in $L_2 (
G_f )$; here the sum ranges over the character space $\Sigma$. For each
character $\chi$, $T_s ( f \ast \chi ) = ( T_s f ) \ast \chi,$ so that $f \ast
\chi$ is almost periodic. The function $f \ast \chi$ is even minimal almost
invariant.

\begin{definition}
  A function $f \in C ( G )$ is called \tmem{left almost invariant} if the
  set $A$ of translations of $f,$ $\Span \{ T_s f : s \in G \}$ is a
  finite-dimensional subspace of $C ( G )$. It is called \textup{almost
  invariant} if it is both left and right almost invariant.
  
  It is called \textup{minimal almost invariant}, if, moreover, every nonzero
  subspace of $A$ which is closed under the translations equals $A.$
\end{definition}

To see that $f \ast \chi$ is almost invariant we recall
from~{\cite{CoquandSpittersPW}} that $\lambda f. \chi \ast f$ is both a
projection and a compact operator, so its range is finite dimensional. To see
that it is minimal we need some preparations.

We consider $C ( G_f )$ as a *-algebra with the convolution operator $\ast$ as
multiplication and the map $\widetilde{}$ defined by $\tilde{f} \assign
\lambda x.f ( x^{- 1} )$ as involution. Then $p$ is called a projection if $p
= p \ast p = \tilde{p} .$

\begin{lemma}
  \label{lem: ideal <less>-<gtr> invariant}
  \textup{{\cite{Naimark/Stern:GroupRep}} (p.216)} A closed subspace in $L_1$
  is a left(right) ideal in $L_1$ if and only if it is invariant under
  left(right) translation. The same holds for subspaces of $L_2 .$
\end{lemma}

\begin{lemma}
  \label{lem: ideal contains projection}Every nonzero closed ideal $I$
  contains a nonzero central element.
\end{lemma}

\begin{proof}
  Remark that $I^{\ast} : = \{ \tilde{f} : f \in I \}$ is a right-ideal, so
  that $I^{\ast} I \subset I^{\ast} \cap I.$ If $f \in I$ and $f \neq 0,$ then
  $\tilde{f} \ast f \neq 0$, because $\tilde{f} \ast f ( e ) = \| f \|_2^2
  \neq 0.$ Hence $I^{\ast} \cap I$ is a closed *-subalgebra which contains a
  nonzero (self-adjoint) element $g$. Since $I$ is both a left and a right
  ideal it is closed under left and right translations, so the projection on
  the center $P_Z f \assign \lambda x. \int f ( zxz^{- 1} ) d z$ is also
  contained in this ideal. 
\end{proof}

Since the only central elements in the ideal $I$ generated by a character
$\chi$ are multiples of this character, we see that any nonzero closed
subideal must actually be equal to $I$. It follows that $f \ast \chi$ is
minimal almost invariant.

This proves the following theorem.

\begin{theorem}
  \label{thm: ap approx L2}Let $f$ be an almost periodic function on a
  topological group $G$. Then $f = \sum f \ast \chi_{}$ in $L_2 ( G_f )$ where
  the sum ranges over $\Sigma$. Moreover, each term $f \ast \chi$ is minimal
  almost invariant.
\end{theorem}

The following theorem is proved in a similar way as Theorem~\ref{Thm: ap unif
R}. It is a Bohr approximation theorem for general topological groups.

\begin{theorem}
  \label{thm: ap approx unif}Let $G$ be a topological group. Every almost
  periodic function on $G$ can be uniformly approximated by a linear
  combination of minimal almost invariant functions which is  almost periodic.
\end{theorem}

\section{Conclusions}

We have given a constructive proof of the Bohr approximation theorem for
general topological groups, thus simplifying and generalizing previous
constructive approaches.

Finally, Loomis~{\cite{Loomis:AHA}} proves that every left almost periodic
function is also right almost periodic. His proof is non-constructive. To be
precise consider his Lemma 41B. Let $n$ be the number of elements of the
family $a_i$ and let $\underline{n}$ denote the finite set with $n$ elements.
Then one needs to isolate all the functions $j : \underline{n} \rightarrow
\underline{n}$ which correspond to a given $b \in G$. In this way we obtain a
subset of a finite approximation to the space of translated functions. Classically,
this approximation is totally bounded, but constructively one needs a finitely
enumerable set. It is unclear to me whether this can be proved constructively.

Parts of this research can already be found in my PhD-thesis~\cite{Spitters:phd}. I would like to thank Wim Veldman for his advice during this period.
Finally, I would like to thank the referees for suggestions that helped to
improve the presentation of the paper.

\bibliographystyle{plain}
\bibliography{ap}

\end{document}